\documentclass[aps, prb, twocolumn, superscriptaddress]{revtex4-1}
\usepackage{amssymb}
\usepackage{graphicx}
\usepackage{dcolumn}
\usepackage{bm}

\begin{document}

\title{Long-lived binary tunneling spectrum in a quantum-Hall
Tomonaga-Luttinger liquid }
\author{K. Washio}
\affiliation{Department of Physics, Tokyo Institute of Technology, 2-12-1 Ookayama,
Meguro, Tokyo, 152-8551, Japan.}
\author{R. Nakazawa}
\affiliation{Department of Physics, Tokyo Institute of Technology, 2-12-1 Ookayama,
Meguro, Tokyo, 152-8551, Japan.}
\author{M. Hashisaka}
\affiliation{Department of Physics, Tokyo Institute of Technology, 2-12-1 Ookayama,
Meguro, Tokyo, 152-8551, Japan.}
\author{K. Muraki}
\affiliation{NTT Basic Research Laboratories, NTT Corporation, 3-1 Morinosato-Wakamiya,
Atsugi, 243-0198, Japan.}
\author{Y. Tokura}
\affiliation{NTT Basic Research Laboratories, NTT Corporation, 3-1 Morinosato-Wakamiya,
Atsugi, 243-0198, Japan.}
\affiliation{Graduate School of Pure and Applied Sciences, University of Tsukuba,
Tsukuba, 305-8571, Japan.}
\affiliation{International Education and Research Center of Science, Tokyo Institute of
Technology, 2-12-1 Ookayama, Meguro, Tokyo, 152-8551, Japan.}
\author{T. Fujisawa}
\email{E-mail: fujisawa@phys.titech.ac.jp}
\affiliation{Department of Physics, Tokyo Institute of Technology, 2-12-1 Ookayama,
Meguro, Tokyo, 152-8551, Japan.}
\date{\today }

\begin{abstract}
The existence of long-lived non-equilibrium states without showing
thermalization, which has previously been demonstrated in time evolution of
ultracold atoms, suggests the possibility of their spatial analogue in
transport behavior of interacting electrons in solid-state systems. Here we
report long-lived non-equilibrium states in one-dimensional edge channels in
the integer quantum Hall regime. An indirect heating scheme in a
counterpropagating configuration is employed to generate a non-trivial
binary spectrum consisting of high- and low-temperature components. This
unusual spectrum is sustained even after travelling 5 - 10 $\mu \mathrm{m}$,
much longer than the length for electronic relaxation (about 0.1 $\mu 
\mathrm{m}$), without showing significant thermalization. This observation
is consistent with the integrable model of Tomonaga-Luttinger liquid. The
long-lived spectrum implies that the system is well described by
non-interacting plasmons, which are attractive for carrying information for
a long distance.
\end{abstract}

\maketitle

\section{Introduction}

Dynamics of quantum many-body systems often result in thermalized states
characterized, for example, by the Fermi distribution function \cite%
{BookGemmer}. Exceptional cases have been discussed for integrable systems,
where an isolated system may be left in a non-equilibrium steady state \cite%
{PolkovnikovRMP2011}. This intriguing aspect has been studied in isolated
systems like one-dimensional (1D) cold atomic chains, where temporal
evolution of the system can be measured after a sudden change of the
interaction (quantum quench) \cite{KinoshitaNature2006,GringScience2012}.
Complementary experiments in transport measurement would allow us to study
spatial evolution of electronic states travelling from a non-interaction
region to an interacting region (spatial analog of quantum quench). The
Tomonaga-Luttinger (TL) liquid, which has been identified in various 1D
wires and quantum-Hall edge channels \cite%
{BookEzawa,ChangRMB,SteinbergNatPhys2008,BarakNatPhys2010,BlumensteinNatPhys2011}%
, is known as an example of integrable systems. Theoretically the Coulomb
interaction between electrons is absorbed under bozonization into
non-interacting plasmon modes of collective density excitations \cite%
{TomonagaPTP1950,LuttingerJMP1963,GiamarchiBook}. Therefore, an ideal TL
liquid never thermalizes as plasmon excitations are conserved during the
transport \cite{CazalillaPRL206,Iucci2009,KennesPRL2013}. Quantum-Hall edge
channels are suitable for demonstrating this non-thermalizing behavior, as
they are well isolated from the environment. Actually, coupling to the
phonon bath is sufficiently weak \cite{JezouinScience2013}, and
backscattering that is unwanted for the TL model is highly forbidden \cite%
{ButtikerPRB1998,vanWeesPRB1989}. Moreover, tunneling spectroscopy with a
quantum dot (QD) allows us to investigate spatial evolution of the
electronic spectrum in a tailored geometry \cite{AltimirasNatPhys2010}.
Previous experiments have identified electronic relaxation from
single-particle excitations to collective excitations (plasmons) in terms of
plasmon boundary scattering known as spin-charge separation\ \cite%
{leSueurPRL2010,BocquillonNatComm2013,InouePRL2014,FreulonNatComm2015} and
fractionalization \cite{KamataNatNano2014,ProkudinaPRL2014} in the TL
physics. However, the absence of thermalization as a hallmark of an
integrable system has not been addressed, as the resulting states were
always close to thermalized states showing a trivial Fermi distribution \cite%
{leSueurPRL2010,DegiovanniPRB2010,KovrizhinPRL2012}.

In this paper we show that quantum-Hall edge channels can actually support
long-lived non-equilibrium states. For this purpose, non-trivial binary
spectrum composed of hot and cold carriers is prepared by an indirect
heating scheme using weakly coupled counterpropagating edge channels.
Quantum dot (QD) spectroscopy clearly reveals that the carriers with the
non-trivial binary spectrum propagate over a long distance, much longer than
the length required for electronic relaxation, without thermalization into a
trivial Fermi distribution. This non-thermalizing characteristics is
consistent with the TL model and encourages us to study non-equilibrium
coherent plasmon transport in the system.

The paper is organized in the following way. After describing an artificial
TL liquid formed in quantum-Hall edge channels (Sec. IIA), we propose a
novel excitation scheme to obtain non-trivial binary spectrum (Sec. IIB).
The binary spectrum is obtained from experiments and simulations. For
experiments, we describe the device structure and measurement scheme (Sec.
IIIA), and compare energy spectra in two regions in distinct geometries
(Secs. IIIB, IIIC, and IIID) by considering the heat flow in the system
(Sec. IIIE). The binary spectrum is found in one of the regions, where small
amplitudes of high-frequency plasmons are excited. For simulations, we
derive plasmon eigenmodes of the system (Sec. IVA), and show that an
approximate binary spectrum appears as a non-equilibrium steady state in the
quantum-quench simulations (Sec. IVB). Finally, we discuss the coupling
strength (Sec VA), the similarities and differences between the experiments
and simulations (Sec. VB), some conserved quantities during the transport
(Sec. VC), and possible relaxation mechanisms in the quantum-Hall system
(Sec. VD) before summarizing the work.

\section{Artificial TL liquid}

\subsection{Plasmon excitation in edge channels}

Figure 1(a) illustrates the interacting quantum-Hall edge channels we study,
where right ($\eta =$ r) and left ($\eta =$ $\ell $) moving electronic
channels with spin $\sigma \in \left\{ \uparrow ,\downarrow \right\} $,
labeled $\left( \eta ,\sigma \right) $, are interacting. Such a geometry can
be prepared by depleting a narrow central region of a two-dimensional
electron system (2DES) in a magnetic field at the filling factor $\nu $ = 2 
\cite{KamataNatNano2014}. In the absence of tunneling and the presence of
electrostatic interaction between the channels, the system mimics an ideal
1D wire described by the standard spin-full TL model \cite%
{GutmanPRL2008,GutmanPRB2009,BergPRL2009,HashisakaPRB2013}. The Coulomb
interaction yields four plasmon eigenmodes, labeled $\left[ d,m\right] $,
i.e., left ($d=$ L) and right ($d=$ R) moving charge ($m=$ C) and spin ($m=$
S) modes, as will be derived in Sec. IVA. The charge and spin modes have
symmetric and anti-symmetric charge distributions, respectively, in the
primary channels (spin-charge separation), where the charge velocity $v_{%
\mathrm{C}}$ is generally greater than the spin velocity $v_{\mathrm{S}}$.
The interaction between counterpropagating channels generates small amount
of dragged charges reverse-travelling in the subsidiary channels\ (charge
fractionalization), as illustrated by coupled wave packets in Fig. 1(a) \cite%
{SafiPRB1995,KamataNatNano2014,ImuraPRB2002}.

\begin{figure}[tbp]
\begin{center}
\includegraphics[width = 3.2in]{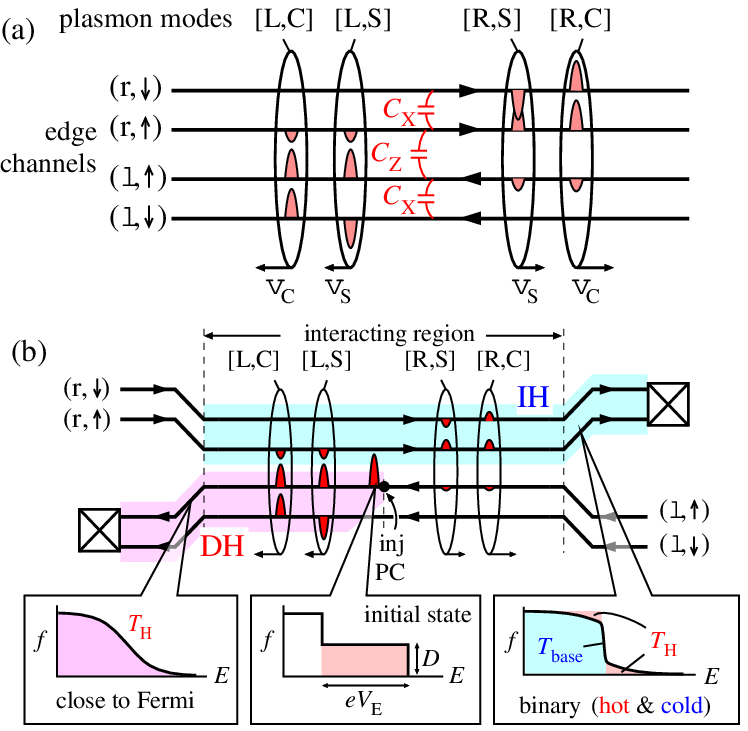}
\end{center}
\caption{ (a) Schematic diagram of interacting quantum Hall edge channels $%
\left( \mathrm{r}/\mathrm{\ell },\uparrow /\downarrow \right) $\ with
nearest neighbor coupling capacitances $C_{\mathrm{X}}$\ and $C_{\mathrm{Z}}$%
. Charge wave packets for plasmon eigenmodes $\left[ \mathrm{L}/\mathrm{R},%
\mathrm{C}/\mathrm{S}\right] $\ are illustrated. (b) The indirect heating
scheme for interacting quantum-Hall edge channels $\left( \mathrm{r}/\mathrm{%
\ell },\uparrow /\downarrow \right) $. Each time a single-electron charge
wave packet is injected from a PC to $\left( \mathrm{\ell },\uparrow \right) 
$ as shown by `inj', four plasmon charge wave packets in the eigenmodes $%
\left[ \mathrm{L}/\mathrm{R},\mathrm{C}/\mathrm{S}\right] $\ are generated.
Carriers are drained to ohmic contacts marked by $\boxtimes $. The energy
distribution function $f\left( E\right) $ shown at the bottom insets
exhibits a double step structure for the initial state (the central inset),
close to Fermi distribution at $T_{\mathrm{H}}$ in the DH region (the left
inset), and an unusual binary spectrum with two temperatures $T_{\mathrm{H}}$
and $T_{\mathrm{base}}$ in the IH region (the right inset).}
\end{figure}

In a practical device, such interacting channels can be formed in a finite
region, outside of which the left and right moving channels are spatially
separated as shown in Fig. 1(b). Suppose that a non-equilibrium charge is
injected into one of the channels, $\left( \ell ,\uparrow \right) $
[illustrated as an arrow marked by `inj'] from a quantum point contact (PC).
The initial electronic state in $\left( \ell ,\uparrow \right) $ exhibits a
double-step profile in the energy distribution function $f\left( E\right) $,
where the width and height are given by the excitation energy $eV_{\mathrm{E}%
}$ and the tunneling probability $D$, respectively, of the PC, as shown in
the central inset at the bottom of Fig. 1(b). However, the interaction
alters the electronic spectrum significantly during the transport \cite%
{leSueurPRL2010}.

In the plasmon picture, charge wave packets randomly injected from the PC
can propagate in a non-interacting manner. Each time a single electron is
injected into $\left( \ell ,\uparrow \right) $, an initial wave packet,
shown just on the left side of the injection point, is generated. It can be
described as a linear superposition of the four plasmon eigenmodes. Since
the charge and spin modes have different velocities, the injected charge
splits into four plasmon packets as illustrated. We will refer to this
splitting as electronic relaxation in the sense that a single-particle
excitation in one channel is relaxed into collective charge excitations over
the channels. This takes place within a length $l_{\mathrm{el}}=\hbar \left(
v_{\mathrm{C}}-v_{\mathrm{S}}\right) /eV_{\mathrm{E}}$, which is
approximately 100 nm at the excitation voltage $V_{\mathrm{E}}$ of 1 mV for
typical velocities $v_{\mathrm{C}}=2-5\times 10^{5}$ m/s and $v_{\mathrm{S}%
}\sim v_{\mathrm{C}}/2$ at $\nu $ = 2 \cite%
{BocquillonNatComm2013,KamataPRB2010,KumadaPRB2011}, from the injection
point. When random charge packets are successively injected from a biased
PC, channels are filled with incoherent ensemble of plasmon wave packets. A
fast wave packet in the charge mode can catch up with a slow packet in the
spin mode, and overtake it without any scattering as the plasmons are
non-interacting. Therefore, the excitation of non-interacting plasmons
cannot be thermalized \cite%
{CazalillaPRL206,KovrizhinPRB2011,LevkivskyiPRB2012}.

\subsection{Non-trivial binary spectrum}

We next show that non-trivial binary spectrum can be generated in the
excitation scheme described in the previous subsection. To see this, it is
important to note that electron-hole excitations (and thus the energy
spectra) in all channels are correlated with each other as they all
originate from the common injection process at the PC. Most of the generated
wave packets flow to the downstream of the injection point, which we refer
to as `direct-heating (DH) region' [the red region in Fig. 1(b)]. Previous
studies have shown that the spectrum in this region can be described by an
approximate Fermi distribution at temperature $T_{\mathrm{H}}$, as
schematically shown in the left inset \cite{leSueurPRL2010}. This might not
contradict the non-thermalizing nature of TL model, as the theory have shown
that the electronic relaxation results in an approximate Fermi distribution
for this case \cite{DegiovanniPRB2010,KovrizhinPRL2012}. However, the
experiment on the DH region was not successful in identifying the presence
of long-lived non-equilibrium states.

Here, we focus on the spectrum in the counterpropagating channels, which we
refer to as `indirect-heating (IH) region' (the blue region). Because of the
small amplitude of the correlated excitations, the spectrum in the IH region
should have a small fraction of the spectrum in the DH region (at $T_{%
\mathrm{H}}$). This fraction $p$ ($\ll 1$) is determined by the interaction,
or the plasmon eigenmodes of the system [See Sec. IV]. The majority of the
spectrum should reflect the base temperature $T_{\mathrm{base}}$ of the
system. This leads to a non-trivial binary spectrum in the IH regions, as
shown in the right inset. The focus of this study is the stability of this
unusual mixture during transport.

This experimental scheme can be understood as a spatial analog of quantum
quench. This analogy is allowed because the plasmon transport is
unidirectional from the injection point through the channels to the ohmic
contacts at the ends of the channels, as is the flow of time in quantum
quench (See discussions in Sec. VB). The initial state prepared as single
electron excitations relaxes into plasmon excitations in the DH and IH
regions. Such spatial evolution can be studied by using PC injectors at
various distances from the QD spectrometer \cite{leSueurPRL2010}. If the
ergodicity is assumed, a many-body state may relax to a thermal-equilibrium
state after travelling a long distance. However, `non-equilibrium steady
state', which can be defined as a state that remains non-equilibrium even
after travelling a sufficiently long (conceptually infinite) distance, may
emerge in an integrable model. The comparison between the experiment in Sec.
III and the simulation in Sec. IV implies that the binary spectrum is the
signature of `non-equilibrium steady state' expected in the integrable model
of TL liquid.

\section{Experiment}

\subsection{Sample and measurement technique}

\begin{figure}[tbp]
\begin{center}
\includegraphics[width = 3.375in]{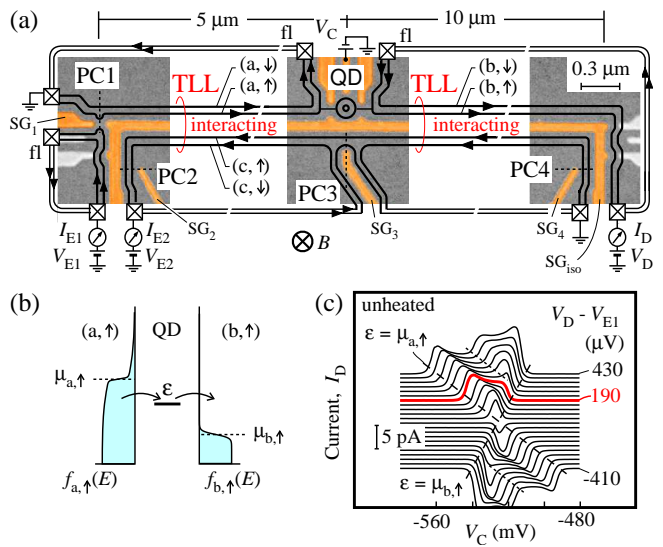}
\end{center}
\caption{ (a) The measurement setup with three magnified
scanning-electron-micrographs with false color. The QD spectrometer and one
of the heat injectors (PC1 - PC4) are activated with the surface gates (SG$_{%
\mathrm{iso}}$, SG$_{\mathrm{1-4}}$, etc.), while inactivated PCs are fully
opened. The double lines indicate edge channels that form when PC1 and PC3
are activated, while only one of the PCs are activated in the actual
measurements. The outer and inner channels serve spin-up and -down
transport, respectively. The QD with a few electrons shows a charging energy
of about 2 meV. Terminals labeled `fl' are floating ohmic contacts. (b) A
schematic energy diagram of the transport through a QD level $\protect%
\varepsilon $. Electronic spectra $f_{\mathrm{a},\uparrow }\left( E\right) $%
\ and $f_{\mathrm{b},\uparrow }\left( E\right) $\ can be determined from the
current $I\left( \protect\varepsilon \right) $. (c) Current spectra of the
QD for various effective bias voltages $V_{\mathrm{D}}-V_{\mathrm{E1}}$\
taken with all PCs (including PC1) fully opened ($G_{i}/G_{\mathrm{q}}=2$).
Dashed lines indicate the energy alignment of the ground state with the
chemical potential of the channels ($\protect\varepsilon =\protect\mu _{%
\mathrm{a}/\mathrm{b},\uparrow }$), while dotted lines indicate that of the
excited states. Each trace is offset for clarity. }
\end{figure}

The measurements were performed with an AlGaAs/GaAs modulation-doped
heterostructure in a perpendicular magnetic field of 5.9 T ($\nu $ = 2) at
the lattice temperature of 20 - 100 mK. The unprocessed 2DES has the
electron density 2.9$\times $10$^{11}$ cm$^{-2}$ and the zero-field mobility
1.6$\times $10$^{6}$ cm$^{2}$/Vs. The device shown in Fig. 2(a) involves
split gates SG's to define the channels, PC charge injectors and a QD
spectrometer. The quantized Hall resistance at $h/2e^{2}$\ and the vanishing
longitudinal resistance of the device ensure the formation of two chiral
edge channels dominating the transport.

\begin{figure*}[t]
\begin{center}
\includegraphics[width = 6.5in]{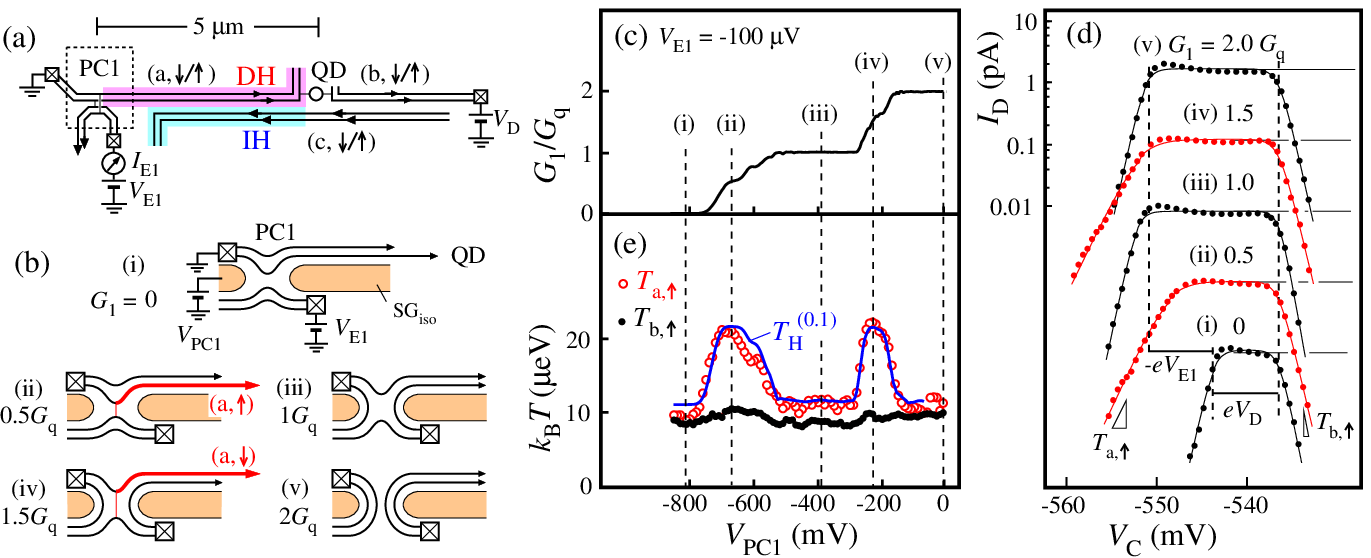}
\end{center}
\caption{ (a) A schematic channel layout for measuring the spectrum in the
DH region formed downstream of PC1 in $\left( \mathrm{a},\uparrow
/\downarrow \right) $. (b) Schematic channel geometries near PC1 for (i) $%
G_{1}$\ = 0, (ii) $G_{1}=0.5G_{\mathrm{q}}$, (iii) $G_{1}=G_{\mathrm{q}}$,
(iv) $G_{1}=1.5G_{\mathrm{q}}$, and (v) $G_{1}=2G_{\mathrm{q}}$. Channels $%
\left( \mathrm{a},\uparrow \right) $\ and $\left( \mathrm{a},\downarrow
\right) $\ can be heated in (ii) and (iv), respectively. (c) $V_{\mathrm{PC1}%
}$\ dependence of $G_{1}/G_{\mathrm{q}}$, measured by the current $I_{%
\mathrm{E1}}$\ at the excitation voltage $V_{\mathrm{E1}}=$\ -100 $\protect%
\mu $V. (d) Current spectra with a flat-top region of the width associated
with the effective bias. Measured current (circles) is fitted well with the
Fermi distribution function (solid lines). The broadened edges measure the
electron temperatures $T_{\mathrm{a/b},\uparrow }$\ in the channels. The $I_{%
\mathrm{D}}$\ and $V_{\mathrm{C}}$\ scales apply to the trace (v), while
other traces are offset for clarity. (e) $k_{\mathrm{B}}T_{\mathrm{a/b}%
,\uparrow }$\ as a function of $V_{\mathrm{PC1}}$. $T_{\mathrm{H}}^{\left(
0.1\right) }$\ is the expected temperature considering the coupling $p$ =
0.1. Small difference between $T_{\mathrm{a},\uparrow }$\ and $T_{\mathrm{b}%
,\uparrow }$\ at unheated conditions (iii) and (v) might come from external
noise in the ammeter for $I_{\mathrm{E}1\text{.}}$ }
\end{figure*}

As shown in Fig. 2(a), three sections of edge channels $\left( \eta ,\sigma
\right) $, where $\eta $ specifies the region `a', `b', and `c', are defined
by using the isolation gate SG$_{\mathrm{iso}}$ (width $w=$ 0.1 $\mu $m) and
a QD spectrometer inserted between the section `a' and `b'. Spin-full TL
liquids labelled TLLs are formed in the interacting regions of $\left( 
\mathrm{a},\sigma \right) $ and $\left( \mathrm{c},\sigma \right) $ for the
length of 5 $\mu $m, as well as $\left( \mathrm{b},\sigma \right) $ and $%
\left( \mathrm{c},\sigma \right) $ of 10 $\mu $m. We investigate how the
spectra of $\left( \mathrm{a},\uparrow \right) $ and $\left( \mathrm{b}%
,\uparrow \right) $ in the vicinity of the QD change when heat current is
injected at different locations with respect to the spectrometer. This is
done by selectively activating one of the four PCs; that is, injection from
PC1 (PC2 - PC4) implies that the spectrometer is placed in the DH (IH)
region(s). The tunneling probability $D_{i,\sigma }$ of $i$-th PC for spin $%
\sigma $ can be estimated from its conductance $G_{i}=G_{\mathrm{q}}\left(
D_{i,\uparrow }+D_{i,\downarrow }\right) $ with $G_{\mathrm{q}}=\frac{e^{2}}{%
h}$, where $G_{1}=I_{\mathrm{E1}}/V_{\mathrm{E1}}$\ for PC1 and $G_{i}=I_{%
\mathrm{E2}}/V_{\mathrm{E2}}$\ for PC$i$\ ($i=$\ 2, 3, and 4) are obtained
from the current $I_{\mathrm{E1/2}}$\ - voltage $V_{\mathrm{E1/2}}$\
characteristics.

Figure 2(c) shows current spectrum with varying effective QD bias $V_{%
\mathrm{D}}-V_{\mathrm{E1}}$ between the drain and the emitter [See Fig.
2(a) for their locations], taken without heating (PC1 fully opend and PC2 -
PC4 fully closed). Current steps associated with the transport through the
ground and excited levels are shown by dashed and dotted lines,
respectively, indicating the level spacing of about 200 $\mu $eV. In the
following experiments, we keep the effective bias voltage below 200 $\mu $V
to probe only the transport through the ground level $\varepsilon $ as shown
in Fig. 2(b).

When $\varepsilon $ is swept with the gate voltage $V_{\mathrm{C}}$, the QD
current measures the electronic spectrum $f_{\mathrm{a}/\mathrm{b},\uparrow
}\left( E\right) $ of channel $\left( \mathrm{a}/\mathrm{b},\uparrow \right) 
$ via the relation 
\begin{equation}
I\left( \varepsilon \right) =I_{0}\left\{ f_{\mathrm{a},\uparrow }\left(
\varepsilon \right) \left[ 1-f_{\mathrm{b},\uparrow }\left( \varepsilon
\right) \right] -f_{\mathrm{b},\uparrow }\left( \varepsilon \right) \left[
1-f_{\mathrm{a},\uparrow }\left( \varepsilon \right) \right] \right\} ,
\label{EqFit}
\end{equation}%
where $I_{0}$ is the saturated current on the step.\ For a large positive
bias, the second term of Eq. \ref{EqFit} describing the negative component
of the current can be neglected. Then, the current profiles around the
onsets $\varepsilon =\mu _{\mathrm{a},\uparrow }$ (the left edge of the peak
at a positive bias) and $\mu _{\mathrm{b},\uparrow }$\ (the right edge)
measure the spectra $f_{\mathrm{a},\uparrow }\left( E\right) $\ and $1-f_{%
\mathrm{b},\uparrow }\left( E\right) $, respectively. Therefore, electronic
spectra $f_{\mathrm{a},\uparrow }\left( E\right) $\ and $f_{\mathrm{b}%
,\uparrow }\left( E\right) $\ can be evaluated from a single current trace $%
I_{\mathrm{D}}\left( V_{\mathrm{C}}\right) $. The unheated spectrum in Fig.
2(c) can be fitted well by using the Fermi distribution 
\begin{equation}
f_{\mathrm{F}}\left( E;T\right) =\left[ e^{\left( E-\mu \right) /k_{\mathrm{B%
}}T}+1\right] ^{-1}  \label{EqFermi}
\end{equation}%
with the thermal energy $k_{\mathrm{B}}T_{\mathrm{a}/\mathrm{b},\uparrow }$
and chemical potential $\mu _{\mathrm{a}/\mathrm{b},\uparrow }$. The sharp
onsets on both sides of the peak indicate the thermal energy $k_{\mathrm{B}%
}T_{\mathrm{base}}=$ 9 - 10 $\mu $eV at the base temperature $T_{\mathrm{base%
}}$ in both channels. The QD spectroscopy relies on transport through a
single level. Therefore we restrict ourselves at low-excitation conditions
where populations to the excited states can be neglected, unless otherwise
stated.

\subsection{Spectrum in the DH region}

First, we investigate the spectrum in the DH region using PC1 located $\sim $
5 $\mu $m upstream of the QD. Figure 3(a) schematically shows the
corresponding channel layout. The DH region is formed in $\left( \mathrm{a}%
,\uparrow \right) $ and $\left( \mathrm{a},\downarrow \right) $ by injecting
heat through PC1 with an excitation voltage $V_{\mathrm{E1}}=$ -100 $\mu $%
eV. Non-equilibrium charge can be injected selectively into $\left( \mathrm{a%
},\uparrow \right) $ for the PC1 conductance $G_{1}$ in the range of $%
0<G_{1}<G_{\mathrm{q}}$ as shown in panel (ii) of Fig. 3(b), and $\left( 
\mathrm{a},\downarrow \right) $ for $G_{\mathrm{q}}<G_{1}<2G_{\mathrm{q}}$
in panel (iv). This was confirmed by measuring clear conductance steps in
the split-gate voltage $V_{\mathrm{PC1}}$ dependence of $G_{1}$ ($=I_{%
\mathrm{E1}}/V_{\mathrm{E1}}$) in Fig. 3(c).

\begin{figure*}[t]
\begin{center}
\includegraphics[width = 6.7in]{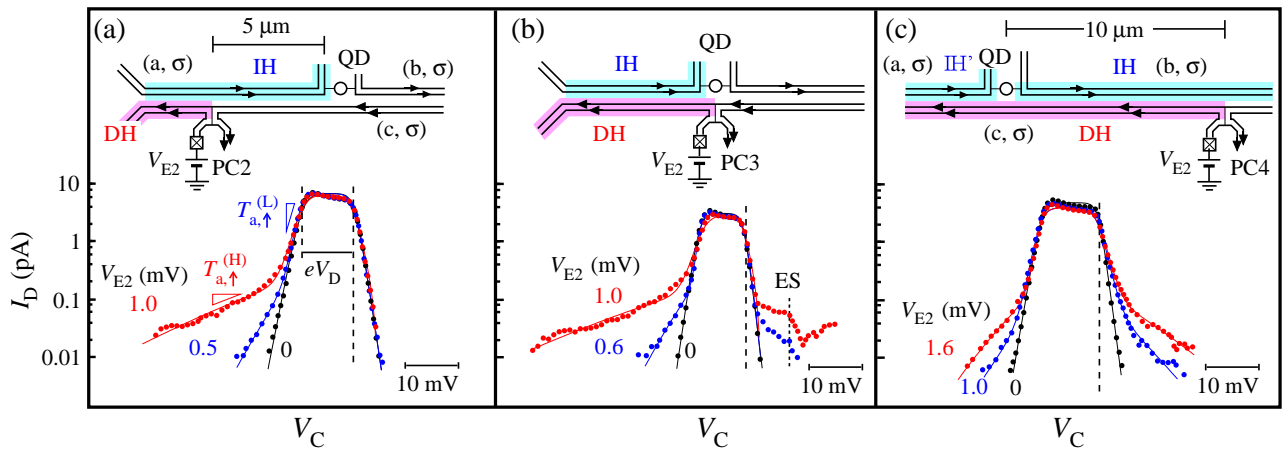}
\end{center}
\caption{ Binary tunneling spectra in the IH regions with heat current
injected from (a) PC2, (b) PC3, and (c) PC4. Measured current (circles) is
fitted well with the binary spectrum (solid lines) with two temperatures, $%
T_{\mathrm{a},\uparrow }^{(\mathrm{H})}$ and $T_{\mathrm{a},\uparrow }^{(%
\mathrm{L})}$, coexisting in the channel. Each inset shows the schematic
channel geometry with the location of IH regions of interest.}
\end{figure*}

Figure 3(d) summarizes the current spectra\ of the QD (filled circles) for
various $G_{1}$ values, plotted as a function of $V_{\mathrm{C}}$. The QD
spectrometer is operated at $V_{\mathrm{D}}=$ 100 $\mu $eV. As the PC1
opening is increased from $G_{1}/G_{\mathrm{q}}=$ $0$ to 2, the width of the
spectrum grows from $eV_{\mathrm{D}}=$ 100 $\mu $eV [trace (i)] to $e\left(
V_{\mathrm{D}}-V_{\mathrm{E1}}\right) =$ 200 $\mu $eV [traces (iii)-(v)],
reflecting the channel $\left( \mathrm{a},\uparrow \right) $ being switched
from the grounded ohmic contact to the one biased at $V_{\mathrm{E1}}$.
Heating in $\left( \mathrm{a},\uparrow \right) $ is manifested in the slope
on the left edge of the spectra,\ which becomes distinctly gentle,
indicating larger $T_{\mathrm{a},\uparrow }$, in the tunneling regimes at $%
G_{1}/G_{\mathrm{q}}=$ $0.5$ [trace (ii)] and 1.5 [trace (iv)]. Note that
heating in $\left( \mathrm{a},\uparrow \right) $ is resolved even when
non-equilibrium charge is injected into $\left( \mathrm{a},\downarrow
\right) $. This ensures the coupling between the two channels. No heating is
seen at quantized conductance $G_{1}/G_{\mathrm{q}}=$ $1$ [trace (iii)] and
2 [trace (v)]. In contrast to the left edge of the spectra, the right edge,
which probes $\left( \mathrm{b},\uparrow \right) $, remains steep with $T_{%
\mathrm{b},\uparrow }\simeq T_{\mathrm{base}}$ for all $G_{1}$ values. This
is reasonable, as no heating is expected for $\left( \mathrm{b},\uparrow
\right) $ in this channel layout [see Fig. 3(a)].

All the spectra in Fig. 3(d) can be fitted well with the Fermi distribution
as shown by the solid lines. Figure 3(e) plots $k_{\mathrm{B}}T_{\mathrm{a}%
,\uparrow }$ (open circles) and $k_{\mathrm{B}}T_{\mathrm{b},\uparrow }$
(solid circles) deduced from the fits, revealing significant heating in $%
\left( \mathrm{a},\uparrow \right) $, for both injection into $\left( 
\mathrm{a},\uparrow \right) $ [$G_{1}/G_{\mathrm{q}}\sim 0.5$ at $V_{\mathrm{%
PC1}}\sim -660$ mV] and $\left( \mathrm{a},\downarrow \right) $ [$G_{1}/G_{%
\mathrm{q}}\sim 1.5$ at $V_{\mathrm{PC1}}\sim -220$ mV]. The fact that $T_{%
\mathrm{a},\uparrow }$ reaches almost the same maximum values at $G_{1}/G_{%
\mathrm{q}}\sim $ 0.5 and 1.5 indicates that the injected heat is equally
partitioned between $\left( \mathrm{a},\uparrow \right) $ and $\left( 
\mathrm{a},\downarrow \right) $. As these data show, the coupling between
hot and cold copropagating channels results in spectra that looks like the
equilibrium Fermi distribution in both of the copropagating channels. Theory
suggests that the actual distribution is different from the Fermi
distribution even in the DH region \cite{DegiovanniPRB2010,KovrizhinPRL2012}%
. However, the deviation is too small to be resolved experimentally,
resulting in seemingly thermalized spectra indistinguishable from the
equilibrium distribution. Consequently, the DH scheme is not suitable for
studying non-equilibrium steady states.

\subsection{Spectrum in the IH region}

In contrast to the case of DH region, qualitatively different tunneling
spectra emerge when the spectrometer is placed in the IH region. IH regions
appear in $\left( \mathrm{a},\sigma \right) $ and/or $\left( \mathrm{b}%
,\sigma \right) $ when one of PC2-PC4 is adjusted in the tunneling regime
with the PC conductance $G_{i}=I_{\mathrm{E2}}/V_{\mathrm{E2}}$ $\sim $ $%
0.5G_{\mathrm{q}}$. Figure 4 summarizes the current spectra (solid circles)
with PC2 in (a), PC3 in (b), and PC4 in (c). As the bias $V_{\mathrm{E2}}$
of the PC is increased, an additional small but long tail develops,
resulting in an anomalous spectrum that cannot be fitted with a single Fermi
distribution function. The long tails appear on one or both sides of the
peak depending on the location of the spectrometer with respect to the PC.

For example, injection from PC2 induces an IH region in $\left( \mathrm{a}%
,\sigma \right) $, but not in $\left( \mathrm{b},\sigma \right) $ [inset of
Fig. 4(a)]. This is consistent with the observation that a tail appears only
on the left side of the peak [anomalous excitation in $\left( \mathrm{a}%
,\uparrow \right) $], by recalling that the current profiles on the left and
right sides of the peak reflect $f_{\mathrm{a},\uparrow }\left( E\right) $\
and $1-f_{\mathrm{b},\uparrow }\left( E\right) $, respectively.

Injection from PC3 should also induce an IH region only in $\left( \mathrm{a}%
,\sigma \right) $. This is consistent with the tail appearing on the left
side [excitation in $\left( \mathrm{a},\uparrow \right) $] as seen in Fig.
4(b). The amplitude of the tail on the left side is comparable to that in
Fig. 4(a). A small step-like profile on the right side is associated with
the excited state of the QD, where electromagnetic and/or phonon-mediated
energy transfer might be responsible only for this geometry with a short
distance between PC3 and QD ($<$\ 0.3 $\mu $m) \cite{OnacPRL2006}. Therefore
this step-like structure is disregarded in the following discussion.

Injection from PC4 induces IH regions in both $\left( \mathrm{a},\sigma
\right) $ and $\left( \mathrm{b},\sigma \right) $, resulting in tails on
both sides [excitation in $\left( \mathrm{a},\uparrow \right) $ and $\left( 
\mathrm{b},\uparrow \right) $] in Fig. 4(c). In this case, PC4 excites left
moving plasmons in modes [L, C/S], which drag non-equilibrium charges in $%
\left( \mathrm{b},\sigma \right) $\ up to the right of the QD. This gives a
tail on the right side of the peak. However, this measurement point is
ill-defined as it is located near the boundary between the IH region and the
unheated upstream region of $\left( \mathrm{b},\sigma \right) $. This could
be the reason of the small amplitude of the right tail as compared to the
tails in Figs. 4(a) and 4(b). Moreover, the left moving plasmons in the
interacting channels $\left( \mathrm{b},\sigma \right) $\ and $\left( 
\mathrm{c},\sigma \right) $\ are interrupted by the QD, and scatter into the
other interacting region with channels $\left( \mathrm{a},\sigma \right) $\
and $\left( \mathrm{c},\sigma \right) $. This plasmon scattering leads
another IH region [denoted by IH' in the inset of Fig. 4(c)] in $\left( 
\mathrm{a},\sigma \right) $, which generates a tail on the left side of the
peak. The small amplitude of the left tail reflects this plasmon scattering.

Apart from the quantitative differences, the unusual tails appear in all of
the IH regions we investigated. As shown by thin solid lines in Fig. 4, all
the spectra in the IH regions can be fitted well with the binary spectrum 
\begin{equation}
f_{\eta ,\sigma }\left( E\right) =\left( 1-p\right) f_{\mathrm{F}}\left(
E;T_{\eta ,\sigma }^{(\mathrm{L})}\right) +pf_{\mathrm{F}}\left( E;T_{\eta
,\sigma }^{(\mathrm{H})}\right) ,  \label{EqBin}
\end{equation}%
which consists of majority carriers at lower temperature $T_{\eta ,\sigma
}^{(\mathrm{L})}$ and minority carriers of a fraction $p$ ($\sim $ 0.1) at
high temperature $T_{\eta ,\sigma }^{(\mathrm{H})}$.

To further demonstrate the validity of Eq. (\ref{EqBin}), we examine the
case of large excitation voltage, where the second term of Eq. \ref{EqFit}
cannot be neglected. We show in Fig. 5 a spectrum excited using PC2 under a
large excitation voltage of $V_{\mathrm{E2}}=$ 2 meV [panel (i)] and compare
it with that without heating [panel (ii)], plotted in a linear scale. As
already shown in Fig. 4(a), in this configuration heating occurs only in $%
\left( \mathrm{a},\sigma \right) $\ near the QD, which affects only the left
side of the peak at low bias. In contrast, the spectrum in panel (i) of Fig.
5 shows a small negative tail (blue hatched region) on the right side, in
addition to the tail on the left side (red hatched region). This signifies
the presence of hot holes as well as hot electrons in $\left( \mathrm{a}%
,\uparrow \right) $. The whole spectrum including the negative tail is
reproduced well with Eq. (\ref{EqBin}) as shown by the solid line. The
distribution functions $f_{\mathrm{a},\uparrow }\left( E\right) $ and $1-f_{%
\mathrm{b},\uparrow }\left( E\right) $ deduced from the fit are shown by
solid lines in the inset,\ highlighting the anomalous binary spectrum $f_{%
\mathrm{a},\uparrow }\left( E\right) $ compared to the Fermi distribution
function $f_{\mathrm{F}}\left( E\right) $ shown by a dashed line.

\begin{figure}[tbp]
\begin{center}
\includegraphics[width = 3.0in]{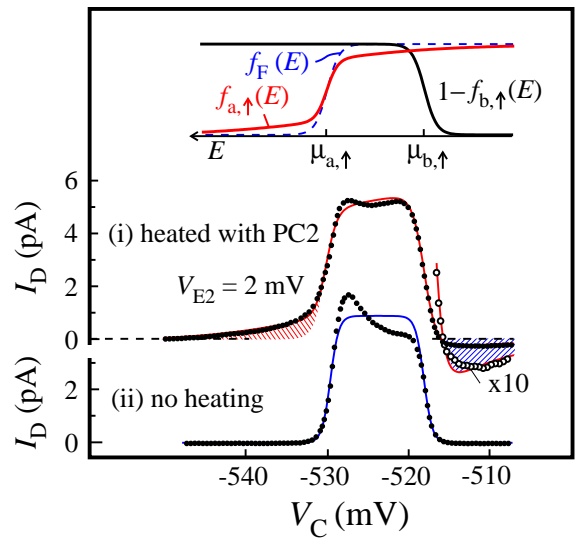}
\end{center}
\caption{ Current spectrum in a linear scale (i) with heat injection from
PC2 and (ii) without heat injection. The positive and negative current tails
in the red and blue hatched regions in (i) represent the excitation of hot
electrons and hot holes, respectively, consistent with the binary spectrum
(the solid lines). The deduced distribution functions $f_{\mathrm{a}%
,\uparrow }\left( E\right) $\ and $1-f_{\mathrm{b},\uparrow }\left( E\right) 
$\ are shown in the inset. The binary spectrum in $f_{\mathrm{a},\uparrow
}\left( E\right) $\ contrasts with a single Fermi distribution function $f_{%
\mathrm{F}}\left( E\right) $\ (dashed line). }
\end{figure}

Binary spectrum can be seen in a wide range of gate voltage, $V_{\mathrm{PC}%
i}$, for $i$-th PC with $i$\ = 2, 3, and 4. Figure 6 summarizes the $V_{%
\mathrm{PC}2}$\ dependence of the PC2 conductance in (a) and the fitting
parameters, the fraction $p$\ in (b) and two distinct temperatures $T_{%
\mathrm{a},\uparrow }^{(\mathrm{H})}$\ (circles) and $T_{\mathrm{a},\uparrow
}^{(\mathrm{L})}$\ (black solid line) in (c),\ for the binary spectrum. At
this large bias voltage $V_{\mathrm{E2}}$\ = 1 mV, $G_{2}$\ in Fig. 6(a)
does not show quantized conductance at $G_{\mathrm{q}}$, which gives
ambiguity in the estimate of $D_{2,\sigma }$. The binary spectrum with high $%
T_{\mathrm{a},\uparrow }^{(\mathrm{H})}$ is seen in the entire tunneling
regime, while $T_{\mathrm{a},\uparrow }^{(\mathrm{L})}$\ as well as $T_{%
\mathrm{b},\uparrow }$\ in the unheated region remain low comparable to the
base temperature.

\begin{figure}[tbp]
\begin{center}
\includegraphics[width = 2.5in]{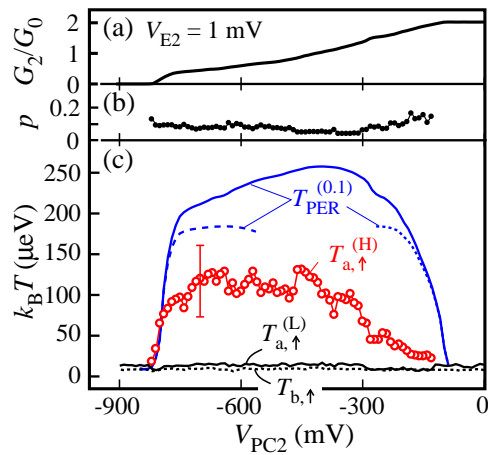}
\end{center}
\caption{(a) Conductance of PC2, $G_{2}=I_{\mathrm{E2}}/V_{\mathrm{E2}}$, as
a function of the gate voltage, $V_{\mathrm{PC2}}$. (b and c) $V_{\mathrm{PC2%
}}$\ dependence of the fraction $p$\ of hot minority carriers in (b) and
thermal energies for $T_{\mathrm{a},\uparrow }^{(\mathrm{H})}$\ (circles), $%
T_{\mathrm{a},\uparrow }^{(\mathrm{L})}$\ (black solid line), and $T_{%
\mathrm{b},\uparrow }$\ (black dashed line) in (c) deduced from the fitting.
The parameter $T_{\mathrm{a},\uparrow }^{(\mathrm{H})}$\ is smaller than but
comparable to $T_{\mathrm{PER}}^{(\mathrm{0.1})}$, which is expected in the
assumption of reaching an identical temperature. The blue solid, dashed, and
dotted lines are $T_{\mathrm{PER}}^{(\mathrm{0.1})}$\ by assuming $%
D_{2,\uparrow }=D_{2,\downarrow }$, $D_{2,\downarrow }=0$, and $%
D_{2,\uparrow }=0$, respectively.}
\end{figure}

Strikingly, the observed binary spectrum did not decay into a single Fermi
function for any geometries we examined. The factor $p$ ($\sim $ 0.1) is
found to be comparable for various distances from the PC to the QD,
including the shortest case with PC3 [Fig. 4(b)], 5 $\mu $m with PC2 [Fig.
4(a)], and 10 $\mu $m with PC4 in the reversed magnetic field (See Appendix
B and Fig. 10). These distances are much longer than $l_{\mathrm{el}}$ ($%
\sim $ 0.1 $\mu $m at $V_{\mathrm{E}}=$ 1 meV) for electronic relaxation to
plasmons. Therefore, the observed binary spectrum supports the picture of
non-interacting plasmon excitations and the non-thermalizing character of
the 1D edge channels.

\subsection{Comparison of the spectra}

The spectra in the DH and IH regions are compared in the same range of
excitation voltages\ in Fig. 7, where the horizontal axis is taken as the
dot energy $\varepsilon $\ relative to $\mu _{\mathrm{a},\uparrow }$. The
spectrum for the DH region at $G_{1}=1.5G_{\mathrm{q}}$\ in Fig. 7(a) is
measured at various $V_{\mathrm{E}1}$\ by keeping the effective bias
constant ($V_{\mathrm{E}1}-V_{\mathrm{D}}=$\ 200 $\mu $V). Heating in $%
\left( \mathrm{a},\uparrow \right) $\ is seen as broadening on the left edge
of the peak. The linear slope over a few orders of magnitude in the
logarithmic scale of $I_{D}$\ suggests that all carriers in the DH region
can be characterized by a single temperature $T^{\left( \mathrm{DH}\right) }$%
. In contrast, the unusual tail for the IH region, which can be recognized
from $V_{\mathrm{E}2}\sim $\ 200 $\mu $V and becomes prominent at $V_{%
\mathrm{E}2}\geq $\ 500 $\mu $V in Fig. 7(b), can be characterized by the
binary spectrum with parameters $T^{\left( \mathrm{H}\right) }$\ (gentle
slope), $T^{\left( \mathrm{L}\right) }$ (steep slope), and $p$.

One can see that $T^{\left( \mathrm{DH}\right) }$\ in the DH region [Fig.
7(a)] is somewhat higher but comparable to $T^{\left( \mathrm{H}\right) }$\
in the IH region [Fig. 7(b)], by comparing the spectra at the same
excitation voltages ($V_{\mathrm{E1}}=V_{\mathrm{E2}}\sim $\ 600 $\mu $V).
This implies the strong relation between the two spectra. Based on the
argument in Sec. IIB, the different spectra in the DH and IH regions can be
understood by considering the amplitudes of the plasmon excitations. The
channels in the DH region are full of hot plasmons with large amplitudes as
shown in Fig. 1(b), which seemingly looks like a thermalized state when the
system is measured with a QD spectrometer. In contrast, the channels in the
IH region involve small amplitudes of hot plasmons in the background of cold
plasmons, which could lead to the anomalous binary spectrum.

We should add some comments on the effect of excited states in the dot. In
Fig. 7(a), a faint kink marked by ES' and a step-like structure marked by ES
appear when the excitation voltage $V_{\mathrm{E}1}$\ exceeds the typical
level spacing (about 200 $\mu $V) of the QD. The former (ES') can be
understood as hot electrons passing through an excited state, and the latter
(ES) can also be attributed to a complex excitation process to an excited
state. Both of these additional features start to appear at $V_{\mathrm{E}1}$%
\ values much higher than that at which the heating in the channel starts to
be visible. In contrast, no features associated with the excited states
appear for the IH region, unless $V_{\mathrm{E}2}$\ exceeds 2 mV. This is
presumably due to the weak coupling ($p$\ $\sim $\ 0.1) between the DH and
IH regions.

\begin{figure}[tbp]
\begin{center}
\includegraphics[width = 3.0in]{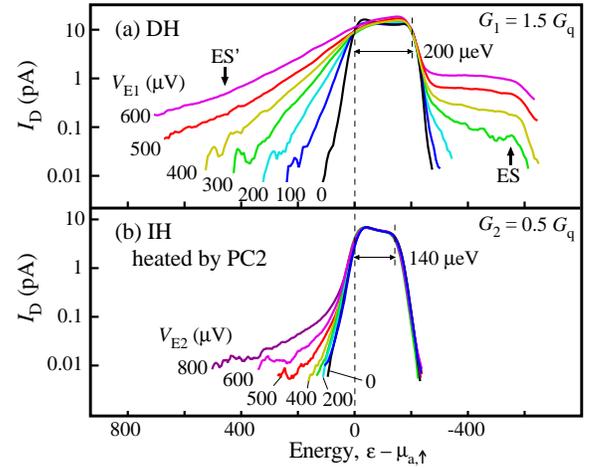}
\end{center}
\caption{(a) Tunneling spectra measured in the DH region induced by PC1 at $%
G_{1}=1.5G_{\mathrm{q}}$. Its $V_{\mathrm{E}1}$\ dependence is taken by
changing both $V_{\mathrm{E}1}$\ and $V_{\mathrm{D}}$\ to keep $V_{\mathrm{E}%
1}-V_{\mathrm{D}}=$\ 200 $\protect\mu $V. Excited-state features are
indicated by arrows (ES and ES'). (b) Tunneling spectra measured in the IH
region induced by PC2 at $G_{2}=0.5G_{\mathrm{q}}$. The binary spectrum
develops with increasing $V_{\mathrm{E}2}$.}
\end{figure}

\subsection{Heat flow}

Provided that the binary distribution of Eq. (\ref{EqBin}) holds well, the
temperatures in DH and IH regions can be related to each other by
considering the heat flow in the system. The heat current injected from the $%
i$-th PC is given by $W_{i}=\frac{1}{2}G_{\mathrm{q}}\sum\nolimits_{\sigma
}D_{i,\sigma }\left( 1-D_{i,\sigma }\right) V_{\mathrm{E}}^{2}$ with the
partitioning factor $D\left( 1-D\right) $ \cite%
{leSueurPRL2010,JezouinScience2013}. This heat should be redistributed over
the four channels. As discussed in Sec IIB, excitations in DH and IH regions
are expected to be correlated. For simplicity, we assume that all carriers
in the DH region and the fraction $p$\ of carriers in the IH region share an
identical temperature $T_{\mathrm{H}}^{(p)}$\ while the cold majority ($1-p$%
) in the IH region remains at $T_{\mathrm{base}}$. This assumption can be
justified for the high-energy spectrum ($\left\vert \varepsilon -\mu _{%
\mathrm{a},\uparrow }\right\vert \gg k_{\mathrm{B}}T_{\mathrm{H}}^{(p)}$) in
the weak coupling limit ($p\ll 1$)\ of quantum quench problem shown in Sec.
IVB. This crude approximation allows us to compare the different experiments
for the DH and IH regions. Then, the energy conservation law suggests that
the spectra are expected to share the same temperature given by $T_{\mathrm{H%
}}^{(p)}=\sqrt{T_{\mathrm{base}}^{2}+\frac{3h}{(1+p)\pi ^{2}k_{\mathrm{B}%
}^{2}}W_{i}}$.

Figure 8(a) shows the bias-voltage $V_{\mathrm{E1}}$ dependence of the
deduced thermal energy $k_{\mathrm{B}}T_{\mathrm{a},\uparrow }$ in the DH
region taken at $G_{2}/G_{\mathrm{q}}\sim $1.5 (circles) and $\sim $0.5
(squares). Figures 8(b) and 8(c) summarize $V_{\mathrm{E2}}$ dependence of
the fitting parameters [$p$\ in (b) and $k_{\mathrm{B}}T_{\mathrm{a}%
,\uparrow }^{(\mathrm{H})}$ (open circles), $k_{\mathrm{B}}T_{\mathrm{a}%
,\uparrow }^{(\mathrm{L})}$ (solid line) and $k_{\mathrm{B}}T_{\mathrm{b}%
,\uparrow }$\ (dotted line) in (c)] for the IH region induced by PC2. Here
we have restricted the ranges $V_{\mathrm{E}1}\leq $\ 0.2 mV and $V_{\mathrm{%
E}2}\leq $\ 1.5 meV, where excited states in the QD play no visible effects
in the spectrum. By choosing a typical value of $p=$ 0.1 from Fig. 8(b), one
can see that $k_{\mathrm{B}}T_{\mathrm{a},\uparrow }$ in Fig. 8(a) as well
as $k_{\mathrm{B}}T_{\mathrm{a},\uparrow }^{(\mathrm{H})}$ in Fig. 8(c)
follow reasonably well the same form of $T_{\mathrm{H}}^{(0.1)}$. This
suggests that the carriers in the DH region and the minority carriers in the
IH region indicate a similar spectrum. Quantitative disagreement, such as
the deviation of $k_{\mathrm{B}}T_{\mathrm{a},\uparrow }^{(\mathrm{H})}$
from a linear dependence in Fig. 8(c), might come from this crude
approximation and possible energy-dependent coupling between the channels 
\cite{ProkudinaPRL2014}. Note that the temperature $T_{\mathrm{a},\uparrow
}^{(\mathrm{L})}$ [solid line in Fig. 8(c)] of the majority carriers always
remains cold. Therefore, the total energy is well conserved in the four
channels with negligible heat leakage to other degrees of freedom, which
ensures that the channels are effectively isolated from the environment. The
same temperature $T_{\mathrm{H}}^{(0.1)}$\ is also plotted in Fig. 6(c),
where one can see that the observed $T_{\mathrm{a},\uparrow }^{(\mathrm{H})}$%
\ is smaller but comparable to $T_{\mathrm{H}}^{(0.1)}$. These evidences
support that the excitations in DH and IH regions are strongly correlated.

\begin{figure}[tbp]
\begin{center}
\includegraphics[width = 3.375in]{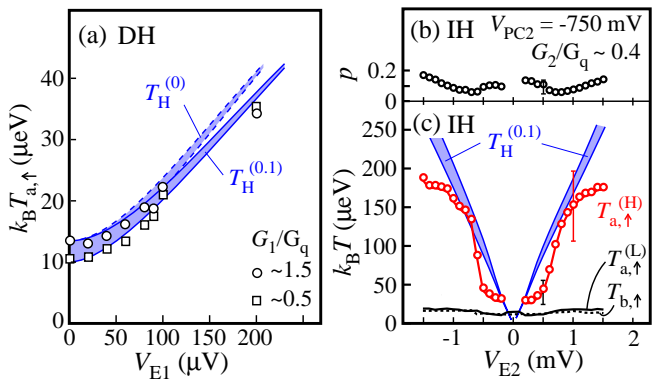}
\end{center}
\caption{ (a) The bias voltage $V_{\mathrm{E1}}$\ dependence of $k_{\mathrm{B%
}}T_{\mathrm{a},\uparrow }$\ obtained at $G_{2}/G_{\mathrm{q}}\sim $1.5
(circles) and $\sim $0.5 (squares). Solid and dashed lines show $T_{\mathrm{H%
}}^{(0.1)}$\ and $T_{\mathrm{H}}^{(0)}$, respectively, considering the
variation of $k_{\mathrm{B}}T_{\mathrm{base}}=$\ 10 (lower bounds) and 13 $%
\protect\mu $eV (upper bounds). (b and c) The bias voltage $V_{\mathrm{E2}}$%
\ dependence of $p$\ in (b) and $k_{\mathrm{B}}T_{\mathrm{a},\uparrow }^{(%
\mathrm{H})}$, $k_{\mathrm{B}}T_{\mathrm{a},\uparrow }^{(\mathrm{L})}$, and $%
k_{\mathrm{B}}T_{\mathrm{b},\uparrow }$\ in (c). The error bars show typical
ambiguity in fitting the binary spectrum. Blue solid lines show $T_{\mathrm{H%
}}^{(0.1)}$, considering the ambiguity of tunneling probabilities between $%
D_{2,\uparrow }=D_{2,\downarrow }=G_{2}/2G_{\mathrm{q}}$\ (upper bound) and $%
D_{2,\uparrow }=G_{2}/G_{\mathrm{q}}$\ and $D_{2,\downarrow }=0$ (lower
bound).}
\end{figure}

Note that the above analysis is based on the assumption where the energy
distribution can be approximated by Fermi distribution in the DH region and
the binary distribution in the IH region. Rigorous analysis should be made
with the TL model, which is beyond the scope of this paper. However,
simulations of quantum quench, a time-dependent problem for initial states
prepared at different temperatures, strongly support this approximation, as
shown in the next chapter.

\section{Simulations with the TL model}

\subsection{\textbf{Plasmon eigenmodes}}

Plasmon modes in the edge channels can be conveniently described by a
capacitance model \cite{KamataNatNano2014,HashisakaPRB2013}. Interaction
inside or between quantum Hall edge channels can be parametrized by
capacitances per unit length, $C_{\mathrm{ch}}$ for intrachannel coupling
and $C_{ij}$ for interchannel coupling between $i$-th and $j$-th channel [$%
i,j\in \left\{ \left( \mathrm{r},\downarrow \right) ,\left( \mathrm{r}%
,\uparrow \right) ,\left( \mathrm{\ell },\uparrow \right) ,\left( \mathrm{%
\ell },\downarrow \right) \right\} $] . For simplicity, we only consider
nearest neighbor couplings with $C_{\mathrm{X}}$ between copropagating
channels and $C_{\mathrm{Z}}$ between counterpropagating channels, as shown
in Fig. 1(a). We assumed that $C_{\mathrm{ch}}$ is identical for all
channels. Excess charge density $\rho _{i}$ and potential $V_{i}$ of $i$-th
channel are related by the coupling capacitances; $\rho _{i}=C_{ij}V_{j}$
with the capacitance matrix of the form%
\begin{equation}
\mathbf{C}=C_{\mathrm{ch}}\left( 
\begin{array}{cccc}
1+c_{\mathrm{X}} & -c_{\mathrm{X}} & 0 & 0 \\ 
-c_{\mathrm{X}} & 1+c_{\mathrm{X}}+c_{\mathrm{Z}} & -c_{\mathrm{Z}} & 0 \\ 
0 & -c_{\mathrm{Z}} & 1+c_{\mathrm{X}}+c_{\mathrm{Z}} & -c_{\mathrm{X}} \\ 
0 & 0 & -c_{\mathrm{X}} & 1+c_{\mathrm{X}}%
\end{array}%
\right)
\end{equation}%
in the order of the channel index $i=\left( \mathrm{r},\downarrow \right)
,\left( \mathrm{r},\uparrow \right) ,\left( \mathrm{\ell },\uparrow \right)
, $ and $\left( \mathrm{\ell },\downarrow \right) $. Here, $c_{\mathrm{X}%
}=C_{\mathrm{X}}/C_{\mathrm{ch}}$ and $c_{\mathrm{Z}}=C_{\mathrm{Z}}/C_{%
\mathrm{ch}}$ denote the normalized coupling strengths. Charge conservation
in each channel can be represented by $\partial \rho _{i}/\partial
t=-S_{i}G_{\mathrm{q}}\partial V_{i}/\partial x$ with the chirality $%
S_{\left( \mathrm{r},\sigma \right) }=1$ for right-moving channels and $%
S_{\left( \ell ,\sigma \right) }=-1$ for left-moving channels, and $G_{%
\mathrm{q}}=\frac{e^{2}}{h}$. Coupled wave equation of the plasmons can be
written in the form 
\begin{equation}
\partial \rho _{i}/\partial t=-M_{ij}\partial \rho _{j}/\partial x
\end{equation}%
with $M_{ij}=S_{i}G_{\mathrm{q}}\left\{ \mathbf{C}^{-1}\right\} _{ij}$.

The plasmon eigenmodes can be obtained by diagonalizing the matrix $\mathbf{M%
}$. The transformation matrix $\mathbf{T}$ that satisfies $\mathbf{T}^{%
\mathbf{\dag }}\mathbf{MT}=\mathrm{diag}(v_{m})$ describes the plasmon
eigenmode $\tilde{\rho}_{m}$ and the velocity $v_{m}$ for $m=\left[ \mathrm{%
R,C}\right] $, $\left[ \mathrm{R,S}\right] $, $\left[ \mathrm{L,S}\right] $,
and $\left[ \mathrm{L,C}\right] $. For a weak coupling limit $c_{\mathrm{Z}%
}\ll c_{\mathrm{X}}\ll 1$, eigenmodes can be approximately written as%
\begin{eqnarray}
\tilde{\rho}_{\mathrm{R,C}} &\simeq &\left( 1,1-c_{\mathrm{Z}}/2c_{\mathrm{X}%
},-c_{\mathrm{Z}}/2,0\right) /\sqrt{2} \\
\tilde{\rho}_{\mathrm{R,S}} &\simeq &\left( -1,1+c_{\mathrm{Z}}/2c_{\mathrm{X%
}},-c_{\mathrm{Z}}/2,0\right) /\sqrt{2} \\
\tilde{\rho}_{\mathrm{L,S}} &\simeq &\left( 0,-c_{\mathrm{Z}}/2,1+c_{\mathrm{%
Z}}/2c_{\mathrm{X}},-1\right) /\sqrt{2} \\
\rho _{\mathrm{L,C}} &\simeq &\left( 0,-c_{\mathrm{Z}}/2,1-c_{\mathrm{Z}%
}/2c_{\mathrm{X}},1\right) /\sqrt{2}
\end{eqnarray}%
with the velocity $v_{\mathrm{R/L,C}}\simeq \pm v_{\mathrm{0}}\left( 1-c_{%
\mathrm{Z}}/2\right) $ for charge modes and $v_{\mathrm{R/L,S}}\simeq \pm v_{%
\mathrm{0}}\left( 1-2c_{\mathrm{X}}-c_{\mathrm{Z}}/2\right) $ for spin
modes, where $v_{\mathrm{0}}=G_{\mathrm{q}}/C_{\mathrm{ch}}$ is the
uncoupled plasmon velocity in a single channel. These plasmon eigenmodes are
schematically shown in Fig. 1(a). Note that these plasmon eigenmodes are
fully consistent with the TL model, as shown in Sec. IVB.

When a unit charge is injected into $\left( \mathrm{\ell },\uparrow \right) $%
, it can be expressed as%
\begin{equation}
\left( 0,0,1,0\right) \simeq \sqrt{2}s\left( \tilde{\rho}_{\mathrm{R,C}}+%
\tilde{\rho}_{\mathrm{R,S}}\right) +\left( \tilde{\rho}_{\mathrm{L,C}}+%
\tilde{\rho}_{\mathrm{L,S}}\right) /\sqrt{2}  \label{Eq0010}
\end{equation}%
with a small fractionalization ratio $s=c_{\mathrm{Z}}/4=C_{\mathrm{Z}}/4C_{%
\mathrm{ch}}$. This explains large and small plasmon amplitudes in the DH
and IH regions, respectively, as schematically illustrated in Fig. 1(b).

The above capacitance model is convenient for describing different coupling
parameters between the channels. If required, the capacitances can be
obtained by solving electrostatic potential around the edge channels \cite%
{KamataNatNano2014}. Actually, the model has been applied to reproduce
plasmon transport in edge channels including plasmon interferometers \cite%
{HashisakaPRB2013}. In this work, we use the matrix $\mathbf{T}$ and
parameters $c_{\mathrm{X}}$ and $c_{\mathrm{Z}}$ to solve the quench problem
shown in the next subsection.

\subsection{Quantum quench}

We have experimentally studied how the energy spectrum changes along the
interacting edge channels under continuously injecting non-equilibrium
electrons from a PC (spatial analog of quantum quench). This can be compared
with the standard quantum quench problem asking how the momentum spectrum
changes with time in the interacting channels after preparing a
non-equilibrium state. Here we apply the quench problem to the
counterpropagating channels to see the emergence of a binary spectrum.

We followed the approach taken by Kovrizhin and Chalker to study the quench
problem for two copropagating channels (see refs. \cite%
{Iucci2009,KovrizhinPRB2011} and references therein). Here, four infinitely
long channels with channel index $j$ (= $1-4$), i.e., two right movers ($j$
= 1 and 2) and two left movers ($j$ = 3 and 4), are initially prepared in
independent thermal equilibrium at different temperatures $T_{j}$ ($%
T_{1}=T_{3}=T_{4}=T_{\mathrm{base}}$ and $T_{2}=T_{\mathrm{H}}>T_{\mathrm{%
base}}$). The initial state of each channel at $t=0$ is described by the
correlation function $\mathcal{G}_{j}\left( x;t=0\right) =\mathcal{G}%
_{T_{j}}^{\left( \mathrm{eq}\right) }\left( x\right) $, where $x$ is
distance and $\mathcal{G}_{T}^{\left( \mathrm{eq}\right) }\left( x\right) =%
\frac{i}{2\hbar }\left( \pi k_{\mathrm{B}}T/\hbar v_{\mathrm{0}}\right)
/\sinh \left[ \left( \pi k_{\mathrm{B}}T/\hbar v_{\mathrm{0}}\right) x\right]
$ is the correlation function in thermal equilibrium at temperature $T$.
Corresponding momentum distribution function $f_{j}\left( k,t=0\right) $ is
the Fermi distribution function $f_{T}^{\left( \mathrm{eq}\right) }\left(
k\right) =1/\left[ \exp \left( \hbar v_{\mathrm{0}}k/k_{\mathrm{B}}T\right)
+1\right] $ at $T=T_{j}$.

Time evolution of the system is calculated based on the standard
bosonization technique for the TL model \cite{vonDelftAnnPhys1998}. The
interaction parameters often described by $g$'s are equivalent to those in
the capacitance model in Sec. IVA; i.e., the so-called $g_{4}$ and $g_{2}$
parameters represent the diagonal and off-diagonal elements, respectively,
of $\mathbf{C}^{-1}$. It should be noted that backscattering with the
so-called $g_{1}$ parameter, which diverts the system away from the TL
model, is well suppressed in the edge channels. The Bogoliubov
transformation can also be performed with the same matrix $\mathbf{T}$.
Then, the dynamics is simply described by non-interacting plasmon modes with
constant velocities. For the independent initial states at $T_{j}$, the
equal-time correlation function $\mathcal{G}_{j}\left( x;t\right) $ at time $%
t$ can be described by a product of initial correlation functions $\mathcal{G%
}_{T_{j}}^{\left( \mathrm{eq}\right) }\left( x\right) $ of all channels as

\begin{eqnarray}
\mathcal{G}_{j}\left( x;t\right) &=&\prod_{j^{\prime }}\left[ \mathcal{G}%
_{T_{j^{\prime }}}^{\left( \mathrm{eq}\right) }\left( x\right) \right]
^{p_{jj^{\prime }}}  \nonumber \\
&&\times \prod_{m,m^{\prime }}\left[ \frac{\mathcal{G}_{T_{j^{\prime
}}}^{\left( \mathrm{eq}\right) }\left( x+\Delta v_{m,m^{\prime }}t\right) }{%
\mathcal{G}_{T_{j^{\prime }}}^{\left( \mathrm{eq}\right) }\left( \Delta
v_{m,m^{\prime }}t\right) }\right] ^{c_{jj^{\prime }m}c_{jj^{\prime
}m^{\prime }}},
\end{eqnarray}%
with the power $p_{jj^{\prime }}=\sum_{m}\left\vert T_{mj}T_{mj^{\prime
}}\right\vert ^{2}$ that satisfies $\sum_{j^{\prime }}p_{jj^{\prime }}=1$. $%
c_{jj^{\prime }m}$ represents the coupling between the channels $j$ and $%
j^{\prime }$ through the plasmon mode $m$, and $\Delta v_{m,m^{\prime }}=$ $%
v_{m^{\prime }}-v_{m^{\prime }}$ is the velocity difference between the
modes $m$ and $m^{\prime }$. For the steady state, $\mathcal{G}_{j}\left(
x;\infty \right) $ at $t\rightarrow \infty $ is given by a power mean of
initial correlation functions as%
\begin{equation}
\mathcal{G}_{j}\left( x;\infty \right) =\prod_{j^{\prime }}\left[ \mathcal{G}%
_{T_{j^{\prime }}}^{\left( \mathrm{eq}\right) }\left( x\right) \right]
^{p_{jj^{\prime }}}.  \label{Gpower}
\end{equation}%
For the weak coupling limit ($c_{\mathrm{Z}}\ll c_{\mathrm{X}}\ll 1$), we
have $p_{jj^{\prime }}\sim 0.5$ between copropagating channels ($%
S_{j}S_{j^{\prime }}=1$) while $p_{jj^{\prime }}$ ($\sim s^{2}$) $\ll 1$
between counterpropagating channels ($S_{j}S_{j^{\prime }}=-1$). Finally,
the momentum distribution function $f_{j}^{\left( \mathrm{st}\right) }\left(
k\right) $ is obtained from the Fourier transform of $\mathcal{G}_{j}\left(
x,\infty \right) $.

When only one channel ($j=2$)\ is heated to $T_{\mathrm{H}}$ with others
kept at $T_{\mathrm{base}}$ in the initial condition, Eq. (\ref{Gpower}) can
be simplified as 
\begin{equation}
\mathcal{G}_{j}\left( x;\infty \right) =\mathcal{G}_{T_{\mathrm{base}%
}}^{\left( \mathrm{eq}\right) }\left[ \mathcal{G}_{T_{\mathrm{H}}}^{\left( 
\mathrm{eq}\right) }/\mathcal{G}_{T_{\mathrm{base}}}^{\left( \mathrm{eq}%
\right) }\right] ^{p_{j2}}.  \label{Gpower2}
\end{equation}%
Focusing on the correlation function at shorter distance ($\left\vert
x\right\vert \lesssim \hbar v_{\mathrm{0}}/k_{\mathrm{B}}T_{\mathrm{H}}$) or
the spectrum at larger momentum ($\left\vert k\right\vert \gtrsim k_{\mathrm{%
B}}T_{\mathrm{H}}/\hbar v_{\mathrm{0}}$), the ratio $X\equiv \mathcal{G}_{T_{%
\mathrm{H}}}^{\left( \mathrm{eq}\right) }/\mathcal{G}_{T_{\mathrm{base}%
}}^{\left( \mathrm{eq}\right) }\sim 1$ to the power of $p\equiv p_{j2}\ll 1$
for $j=3$ or 4 can be approximated to $X^{p}\simeq 1+p\left( X-1\right)
+O\left( \left( X-1\right) ^{2}\right) $. This yields $\mathcal{G}_{j}\simeq
\left( 1-p\right) \mathcal{G}_{T_{\mathrm{base}}}^{\left( \mathrm{eq}\right)
}+p\mathcal{G}_{T_{\mathrm{H}}}^{\left( \mathrm{eq}\right) }$. Its Fourier
spectrum exhibits a binary spectrum of the form%
\begin{equation}
f_{j=3,4}^{\left( \mathrm{st}\right) }\left( k\right) \simeq \left(
1-p\right) f_{T_{\mathrm{base}}}^{\left( \mathrm{eq}\right) }\left( k\right)
+pf_{T_{\mathrm{H}}}^{\left( \mathrm{eq}\right) }\left( k\right)
\end{equation}%
in the counterpropagating channels ($j=$ 3 or 4). In this way, the
appearance of an approximate binary spectrum is suggested in the quantum
quench problem. In contrast, electronic spectrum in the copropagating
channels ($j=1$\ and 2) can be approximated to $f_{j=1,2}^{\left( \mathrm{st}%
\right) }\left( k\right) \simeq f_{T_{\mathrm{H}}}^{\left( \mathrm{eq}%
\right) }\left( k\right) $\ by neglecting an anomaly around $k=0$.
Therefore, the two spectra in counter- and copropagating channels share an
identical spectrum. This characteristics is used in the analysis of heat
flow in Sec. IIIE.

The exact solution of the spectra are neither the binary spectrum nor the
Fermi distribution function. The accurate spectrum can be obtained by
performing a Fourier transform of Eq. (\ref{Gpower2}). The solid lines in
Fig. 9(a) and 9(b) show the initial momentum distribution functions $%
f_{j}^{\left( \mathrm{ini}\right) }\left( k\right) $ and the steady-state
distribution function $f_{j}^{\left( \mathrm{st}\right) }$, respectively,
plotted as a function of the normalized momentum $\hbar v_{\mathrm{0}}k/k_{%
\mathrm{B}}T_{\mathrm{base}}$. Here, we assumed $T_{\mathrm{H}}=10T_{\mathrm{%
base}}$ and coupling parameters $c_{\mathrm{X}}=0.8$ and $c_{\mathrm{Z}}=0.6$%
. The parameters, which are realistic in the capacitance model in Sec. IVA,
were chosen to imitate the binary spectra in Fig. 4. The copropagating
channels 1 and 2 resulted in similar spectra ($f_{1}\sim f_{2}$), which
resemble a single Fermi distribution function shown by the dashed line. In
contrast, anomalous spectra ($f_{3}$ and $f_{4}$) emerge in the
counterpropagating channels 3 and 4. The spectrum can be approximated by a
binary spectrum composed of hot and cold carriers, as shown by the
dash-dotted line. In this range of plots in Fig. 9(b), the temperature (the
inverse of the slope) of hot minority carriers in the counterpropagating
channels is somewhat lower than that in copropagating channels.

In this way, the unusual binary spectrum can emerge in quantum-quench
simulation. The simulation can be compared to the experiment by regarding
the heated channel ($j=2$) as $\left( \ell ,\uparrow \right) $ in Fig. 1(b),
spectra $f_{1}$ and $f_{2}$ as those in the DH region, and spectra $f_{3}$
and $f_{4}$ as those in the IH region.

\begin{figure}[tbp]
\begin{center}
\includegraphics[width = 3.2in]{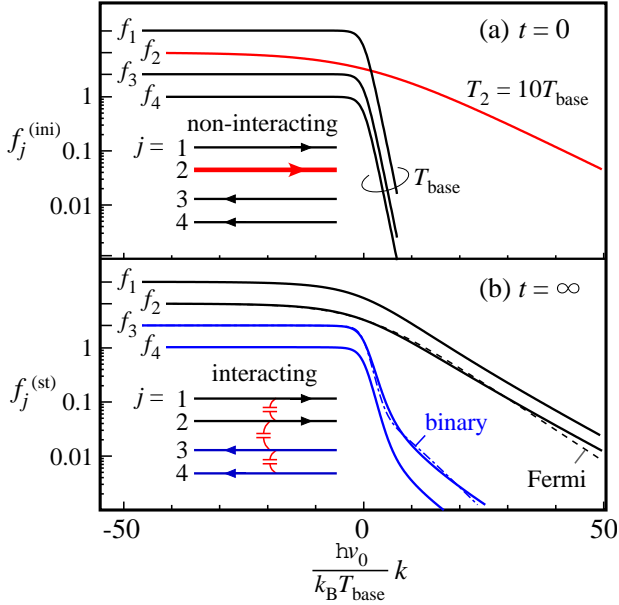}
\end{center}
\caption{ Momentum distribution functions for (a) initial states in
non-interacting channels and (b) steady state at $t=\infty $ after the
interaction is turned on at $t=0$. Each trace is offset for clarity.
Initially, channels are in thermal equilibrium at $T_{1}=T_{3}=T_{4}=T_{%
\mathrm{base}}$ and $T_{2}=$ 10$T_{\mathrm{base}}$ in (a). Non-equilibrium
steady states are reached at $t=\infty $ in (b), where quasi-binary spectrum
emerges in counter-propagating channels 3 and 4. The dashed and dash-dotted
lines show single and binary Fermi distribution functions, respectively,
adjusted to the spectra based on the TL theory (solid lines). The schematic
channel geometries without and with interaction are shown in the inset of
(a) and (b), respectively.}
\end{figure}

\section{Discussion}

\subsection{Coupling strength}

The factor $p$\ deduced from the fits, which ranges from 0.03 to 0.2 in our
PCs, measures the coupling between the counterpropagating channels across
the isolation gate ($G_{\mathrm{iso}}$) of the width \ $w$\ $\sim $\ 0.1 $%
\mu $m. This can be compared with those in previous studies with
non-spectroscopic means. The time-domain charge measurement in Ref. \cite%
{KamataNatNano2014} has demonstrated charge fractionalization of an incident
wave packet, which yields the fractionalization ratio of $r=$ 0.04 in the
amplitude, for the gate width $w$\ $\sim $\ 1 $\mu $m. Based on the model
shown in Sec. IV, this corresponds to $p\sim 4r^{2}\sim $\ 0.006 in our
parameter. Bolometric detection of heat transfer between counterpropagating
channels in Ref. \cite{ProkudinaPRL2014} have identified the interaction
strength $\left\vert 1-K\right\vert =$\ 0.1 $\sim $\ 0.25, which corresponds
to $p\sim 4\left( 1-K^{-1}\right) ^{2}=$\ 0.03 $\sim $\ 0.16, depending on
the gate voltage on the isolation gate of width $w$\ $\sim $\ 0.2 $\mu $m.
Since the fraction $p$\ should increase with decreasing the separation of
the counterpropagating channels, $p$\ $\sim $\ 0.1 in our case is reasonable
for $w$\ $\sim $\ 0.1 $\mu $m.

\subsection{Analogy with quantum quench}

The time dependence of the momentum spectrum (quantum quench) in Sec. IV can
be related to the spatial dependence of the energy spectrum studied in Sec.
III. The differences and similarities between them are summarized as follows.

Firstly, the energy (momentum) spectrum is given by the Fourier transform of
the equal-position (equal-time) correlation function. Considering the
non-interacting non-dispersive plasmon modes with the constant velocities,
the energy spectrum and the momentum spectrum should be related to each
other \cite{KovrizhinPRB2011,GutmanPRB2009,JurcevicNature2014}.

Secondly, the initial state that can be prepared by a PC shows a double-step
distribution function [see the central inset of Fig. 1(b)], which is not in
thermal equilibrium assumed in the quantum-quench simulation. Previous
experiments for two copropagating channels \cite{leSueurPRL2010}, as well as
our results in Sec. IIIB, resulted in a spectrum close to Fermi distribution
function, which is similar to the spectra $f_{1}$ and $f_{2}$ in Fig. 9(b).
This behavior can be reproduced in simulations of quantum quench starting
from a double-step function in one channel \cite{KovrizhinPRL2012} as well
as from thermal equilibrium states at different temperatures \cite%
{KovrizhinPRB2011}. This suggests that these different initial states does
not play significant roles in the final state for the problem. Therefore,
quantum quench starting from a Fermi distribution function is not a bad
assumption to validate the appearance of binary spectrum in the system.

Thirdly, different from the quantum quench problem, the length of the
interacting region in our experiment is finite. Thanks to the chiral nature
of quantum-Hall\ edge channels, the plasmon transport is unidirectional
particularly outside the interacting region. Non-equilibrium charge that
left the interacting region never returns back. This prevents unwanted
effects in the leads and dissipative ohmic contacts, and allows us to
neglect the finite-length problem. Such unidirectional transport cannot be
expected for conventional 1D wires connected to diffusive leads.

In this way, our experiment can be regarded as spatial analog of quantum
quench.

\subsection{Conserved quantities during the transport}

Non-equilibrium transport in integer quantum Hall edge channels has been
investigated for a few decades. When the scattering between the channels is
well suppressed, the chemical potential of each channel is conserved during
the transport, allowing the system to have independent charge distributions
on the channels \cite{ButtikerPRB1998,vanWeesPRB1989}. When the coupling to
the environment, such as phonon bath, is negligible, the heat or the
electron temperature of the system is conserved during the transport \cite%
{GrangerPRL2009,JezouinScience2013,ProkudinaPRL2014}. More intriguingly,
when the interaction is not fully ergodic to cause thermal equilibration,
the system may have infinite numbers of conserved quantities \cite%
{BergesPRL2004}. Our spectroscopic analysis has successfully revealed
non-equilibrium electronic distribution with, at least, four conserved
quantities (the binary spectrum parametrized by $T^{(\mathrm{H})}$, $T^{(%
\mathrm{L})}$, $p$, as well as $\mu $), which is greater than two ($\mu $
and $T$) for trivial cases.

The TL theory suggests that the plasmons are non-interacting and thus should
be conserved during the transport. This is supported by the fact that the
non-equilibrium spectrum is sustained for a long distance. This is
attractive for carrying much information as plasmon excitations rather than
short-lived electronic excitations \cite{JiNature2003,LevkivskyiPRB2008}.

\subsection{Possible relaxation mechanisms}

In previous reports, other energy transfer mechanisms such as phonons \cite%
{ProkudinaPRB2010} or impurities \cite{LundePRB2010} have been discussed to
explain the charge and heat transfer. Such extrinsic effects with random
processes are expected to result in a continuous thermalization into a
trivial Fermi distribution. This contrasts with our binary spectrum
independent of the propagation length (from $\sim $0 to 10 $\mu $m), which
implies that the extrinsic thermalization processes becomes relevant only at
longer distances. This signifies the importance of Coulomb interaction in 1D
systems prepared in the integer quantum Hall regime.

\section{Summary}

We have investigated the energy spectrum of non-equilibrium states in
quantum-Hall Tomonaga-Luttinger liquids. When non-equilibrium charge is
injected from a PC, a non-trivial binary spectrum consisting of high- and
low-temperature components appears in the IH region while a seemingly
thermalized spectrum appears in the DH region. The binary spectrum is
sustained even after travelling 5 - 10 $\mu \mathrm{m}$, much longer than
the length for electronic relaxation (about 0.1 $\mu \mathrm{m}$), without
showing significant thermalization. This can be compared with the simulation
of quantum quench problem, which also suggest the emergence of an
approximate binary spectrum as a non-equilibrium steady state. The
long-lived binary spectrum implies that the system is well described by
non-interacting plasmons, which suggests that low-energy excitations of edge
channels in the integer quantum Hall regime can be well understood as a TL
liquid.

\begin{figure}[tb]
\begin{center}
\includegraphics[width = 3.375in]{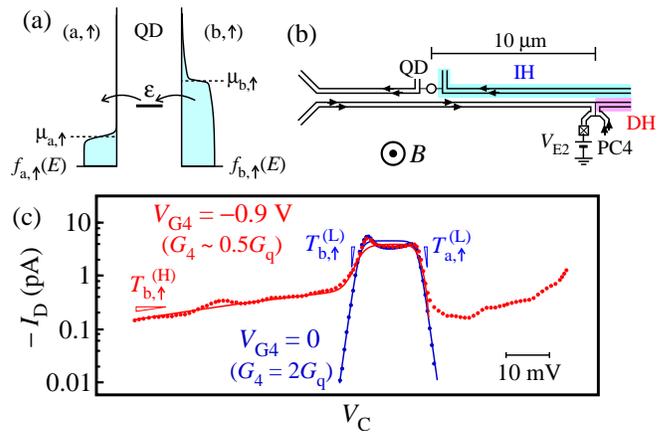}
\end{center}
\caption{ Spectrum after travelling 10 $\protect\mu $m. (a) The energy
diagram at a negative bias $\protect\mu _{\mathrm{a},\uparrow }-\protect\mu %
_{\mathrm{b},\uparrow }=eV_{\mathrm{D}}$ = -80 $\protect\mu $eV. (b) The
location of DH and IH regions in the schematic channel layout. (c) Current
spectrum of the QD measured with PC4 in the reversed magnetic field. Binary
spectrum is seen on the left side of the peak only when the PC4 conductance
is set in the tunneling regime ($G_{4}\simeq 0.5G_{\mathrm{q}}$). }
\end{figure}

\begin{acknowledgments}
We would like to thank Tsuneya Ando, Norio Kawakami, Tomohiro Sasamoto, and
Masahito Ueda for fruitful discussions. This work was supported by Japan
Society for the Promotion of Science (JSPS) KAKENHI (Grant Numbers 21000004,
12J09291, 24009291, 26247051, and 15H05854), International Research Center
for Nanoscience and Quantum Physics at Tokyo Institute of Technology, and
Nanotechnology Platform Program of the Ministry of
Education,Culture,Sports,Science and Technology, Japan.
\end{acknowledgments}

\appendix

\section{Fitting procedure}

The fitting curves in Figs. 3, 4 and 5 are obtained by using Eq. (\ref{EqFit}%
) with a constant current level $I_{0}$ as a free parameter. The dot energy $%
\varepsilon =\alpha \left( V_{\mathrm{C}}-V_{\mathrm{C,0}}\right) $ is
converted from the gate voltage $V_{\mathrm{C}}$ with a factor $\alpha
\simeq $ 0.018$e$\ from an offset $V_{\mathrm{C,0}}$. For the spectrum in
the DH region in Fig. 3(b), the Fermi distribution functions [Eq. (\ref%
{EqFermi})] with free parameters $T_{\mathrm{a}/\mathrm{b},\uparrow }$ and $%
\mu _{\mathrm{a}/\mathrm{b},\uparrow }$ are used. When binary spectrum is
found in the IH regions (Figs. 4 and 5), the binary spectra of the form Eq. (%
\ref{EqBin}) with free parameters $p_{\mathrm{a}/\mathrm{b},\uparrow }$, $T_{%
\mathrm{a}/\mathrm{b},\uparrow }^{(\mathrm{L})}$, $T_{\mathrm{a}/\mathrm{b}%
,\uparrow }^{(\mathrm{H})}$, and $\mu _{\mathrm{a}/\mathrm{b},\uparrow }$
are used. We fitted $I_{\mathrm{fit}}\left( \varepsilon \right) $ to the
measured data $I_{\mathrm{meas}}\left( \varepsilon \right) $ by minimizing
the residual error $\int \left[ \log I_{\mathrm{meas}}\left( \varepsilon
\right) -\log I_{\mathrm{fit}}\left( \varepsilon \right) \right]
^{2}d\varepsilon $ evaluated in the logarithmic scale to focus on the
low-current profile. In order to obtain reliable fitting, data with current
level smaller than 10 - 20 fA were discarded. The hot minority spectrum with
wide current level ranging from 10 fA - 200 fA is analyzed. Since small
fitting error always exists near the transition between the hot and cold
spectrum, the fitting involves ambiguity as shown by error bars in Figs.
6(c) and 8(c). When $T^{(\mathrm{H})}$ is high, the spectrum is extended
over a wide energy range greater than the typical energy spacing of 200 $\mu 
$eV. We neglected any effects from the excited states and energy-dependent
tunneling rate, which might give additional error in the estimate of $T^{(%
\mathrm{H})}$.

\section{Binary spectrum after travelling 10 $\protect\mu $m}

The spectrum measured at the farthest position downstream in the IH region
is obtained with PC4 in the reversed field at $B$ = -5.9 T. As shown in Fig.
10(b), QD is located 10 $\mu $m downstream in the IH region from PC4. As
shown in Fig. 10(c), a binary spectrum emerges on the left side of the peak
only when a large excitation voltage ($V_{\mathrm{E2}}$ = 2 mV) is applied
across PC4 in the tunneling regime ($G_{4}\simeq $ 0.5$G_{\mathrm{q}}$ at $%
V_{\mathrm{G4}}$ = -0.9 V). The profile can be fitted with the binary
spectrum of hot minority carriers ($k_{\mathrm{B}}T_{\mathrm{b},\uparrow }^{(%
\mathrm{H})}\simeq $ 350 $\mu $eV and $p\simeq $ 0.25) and cold majority
carriers ($k_{\mathrm{B}}T_{\mathrm{b},\uparrow }^{(\mathrm{L})}\simeq $ 15 $%
\mu $eV) as shown by the thin solid line. The large current on the right of
the peak is associated with a similar binary spectrum detected with the next
Coulomb blockade peak. Note that, for this experiment, negative bias $V_{%
\mathrm{D}}$ = -80 $\mu $V is applied to the channel $\left( \mathrm{b}%
,\sigma \right) $ to make the transport go from the right to the left as
shown in Fig. 10(a). Therefore, the spectrum of $\left( \mathrm{b},\uparrow
\right) $ appears on the left side of the peak, which is different from the
data set in Fig. 4. 


\begin{thebibliography}{99}
\bibitem{BookGemmer} J. Gemmer, M. Michel, and G. Mahler, \textit{Quantum
Thermodynamics} (Lecture Notes in Physics 784, Springer, 2006).

\bibitem{PolkovnikovRMP2011} A. Polkovnikov, K. Sengupta, A. Silva, and M.
Vengalattore, \textit{Rev. Mod. Phys.} \textbf{83}, 863 (2011).

\bibitem{KinoshitaNature2006} T. Kinoshita, T. Wenger, and D. S. Weiss, 
\textit{Nature} \textbf{440}, 900 (2006).

\bibitem{GringScience2012} M. Gring, M. Kuhnert, T. Langen, T. Kitagawa, B.
Rauer, M. Schreitl, I. Mazets, D. A. Smith, E. Demler, and J. Schmiedmayer, 
\textit{Science} \textbf{337}, 1318 (2012).

\bibitem{BookEzawa} Z. F. Ezawa, \textit{Quantum Hall Effects: Field
Theoretical Approach and Related Topics} (World Scientific, 2008).

\bibitem{ChangRMB} A. M. Chang, \textit{Rev. Mod. Phys.} \textbf{75}, 1449
(2003).

\bibitem{SteinbergNatPhys2008} H. Steinberg, G. Barak, A. Yacoby, L. N.
Pfeiffer, K. W. West, B. I. Halperin, and K. Le Hur, \textit{Nature Phys.} 
\textbf{4}, 116 (2008).

\bibitem{BarakNatPhys2010} G. Barak, H. Steinberg, L. N. Pfeiffer, K. W.
West, L. Glazman, F. von Oppen, and A. Yacoby, \textit{Nat. Phys.} 6, 489
(2010).

\bibitem{BlumensteinNatPhys2011} C. Blumenstein, J. Schafer, S. Mietke, S.
Meyer, A. Dollinger, M. Lochner, X. Y. Cui, L. Patthey, R. Matzdorf, and R.
Claessen, \textit{Nat. Phys.} 7, 776 (2011).

\bibitem{TomonagaPTP1950} S. Tomonaga, \textit{Prog. Theor. Phys.} \textbf{5}%
, 544 (1950).

\bibitem{LuttingerJMP1963} J. M. Luttinger, \textit{J. Math. Phys.} \textbf{4%
}, 1154 (1963).

\bibitem{GiamarchiBook} T. Giamarchi, \textit{Quantum Physics in One
Dimension} (Oxford Univ. Oxford, 2004).

\bibitem{CazalillaPRL206} M. A. Cazalilla, \textit{Phys. Rev. Lett.} \textbf{%
97}, 156403 (2006).

\bibitem{Iucci2009} A. Iucci and M. A. Cazalilla, \textit{Phys. Rev. A} 
\textbf{80}, 063619 (2009).

\bibitem{KennesPRL2013} D. M. Kennes, C. Kl\"{o}ckner, and V. Meden, \textit{%
Phys. Rev. Lett.} \textbf{113}, 116401 (2014).

\bibitem{JezouinScience2013} S. Jezouin, F. D. Parmentier, A. Anthore, U.
Gennser, A. Cavanna, Y. Jin, and F. Pierre, \textit{Science} \textbf{342},
601 (2013).

\bibitem{ButtikerPRB1998} M. B\"{u}ttiker, \textit{Phys. Rev. B} \textbf{38}%
, 9375 (1988).

\bibitem{vanWeesPRB1989} B. J. van Wees, E. M. M. Willems, L. P.
Kouwenhoven, C. J. P. M. Harmans, J. G. Williamson, C. T. Foxon, and J. J.
Harris, \textit{Phys. Rev. B} \textbf{39}, 8066 (1989).

\bibitem{AltimirasNatPhys2010} C. Altimiras, H. le Sueur, U. Gennser, A.
Cavanna, D. Mailly, and F. Pierre, \textit{Nature Phys.} \textbf{6}, 34
(2010).

\bibitem{leSueurPRL2010} H. le Sueur, C. Altimiras, U. Gennser, A. Cavanna,
D. Mailly, and F. Pierre, \textit{Phys. Rev. Lett.} \textbf{105}, 056803
(2010).

\bibitem{BocquillonNatComm2013} E. Bocquillon, V. Freulon, J. M. Berroir, P.
Degiovanni, B. Pla\c{c}ais, A. Cavanna, Y. Jin, and G. F\`{e}ve, \textit{%
Nature Commun.} \textbf{4}, 1839 (2013).

\bibitem{InouePRL2014} H. Inoue, A. Grivnin, N. Ofek, I. Neder, M. Heiblum,
V. Umansky, and D. Mahalu, \textit{Phys. Rev. Lett.} \textbf{112}, 166801
(2014).

\bibitem{FreulonNatComm2015} V. Freulon, A. Marguerite, J. M. Berroir, B.
Placais, A. Cavanna, Y. Jin, and G. Feve, \textit{Nat Commun.} \textbf{6},
6854 (2015).

\bibitem{KamataNatNano2014} H. Kamata, N. Kumada, M. Hashisaka, K. Muraki,
and T. Fujisawa, \textit{Nature Nanotechnol.} \textbf{9}, 177 (2014).

\bibitem{ProkudinaPRL2014} M. G. Prokudina, S. Ludwig, V. Pellegrini, L.
Sorba, G. Biasiol, and V. S. Khrapai, \textit{Phys. Rev. Lett.} \textbf{112}%
, 216402 (2014).

\bibitem{DegiovanniPRB2010} P. Degiovanni, C. Grenier, G. F\`{e}ve, C.
Altimiras, H. le Sueur, and F. Pierre, \textit{Phys. Rev. B} \textbf{81},
121302 (2010).

\bibitem{KovrizhinPRL2012} D. L. Kovrizhin and J. T. Chalker, \textit{Phys.
Rev. Lett.} \textbf{109}, 106403 (2012).

\bibitem{GutmanPRL2008} D. B. Gutman, Y. Gefen, and A. D. Mirlin, \textit{%
Phys. Rev. Lett.} \textbf{101}, 126802 (2008).

\bibitem{GutmanPRB2009} D. B. Gutman, Y. Gefen, and A. D. Mirlin, \textit{%
Phys. Rev. B} \textbf{80}, 045106 (2009).

\bibitem{BergPRL2009} E. Berg, Y. Oreg, E.-A. Kim, and F. von Oppen, \textit{%
Phys. Rev. Lett.} \textbf{102}, 236402 (2009).

\bibitem{HashisakaPRB2013} M. Hashisaka, H. Kamata, N. Kumada, K. Washio, R.
Murata, K. Muraki, and T. Fujisawa, \textit{Phys. Rev. B} \textbf{88},
235409 (2013).

\bibitem{SafiPRB1995} I. Safi and H. J. Schulz, \textit{Phys Rev B} \textbf{%
52}, R17040 (1995).

\bibitem{ImuraPRB2002} K. I. Imura, K. V. Pham, P. Lederer, and F. Piechon, 
\textit{Phys. Rev. B} \textbf{66}, 035313 (2002).

\bibitem{KamataPRB2010} H. Kamata, T. Ota, K. Muraki, and T. Fujisawa, 
\textit{Phys. Rev. B} \textbf{81}, 085329 (2010).

\bibitem{KumadaPRB2011} N. Kumada, H. Kamata, and T. Fujisawa, \textit{Phys.
Rev. B} \textbf{84}, 045314 (2011).

\bibitem{KovrizhinPRB2011} D. L. Kovrizhin and J. T. Chalker, \textit{Phys.
Rev. B} \textbf{84}, 085105 (2011).

\bibitem{LevkivskyiPRB2012} I. P. Levkivskyi and E. V. Sukhorukov, \textit{%
Phys. Rev. B} \textbf{85}, 075309 (2012).

\bibitem{OnacPRL2006} E. Onac, F. Balestro, L. H. W. van Beveren, U.
Hartmann, Yu. V. Nazarov, and L. P. Kouwenhoven, \textit{Phys. Rev. Lett. }%
\textbf{96}, 176601 (2006).

\bibitem{vonDelftAnnPhys1998} J. von Delft and H. Schoeller, \textit{Ann.
Phys.} \textbf{7}, 225 (1998).

\bibitem{JurcevicNature2014} P. Jurcevic, B. P. Lanyon, P. Hauke, C. Hempel,
P. Zoller, R. Blatt, and C. F. Roos, \textit{Nature} \textbf{511}, 202
(2014).

\bibitem{GrangerPRL2009} G. Granger, J. P. Eisenstein, and J. L. Reno, 
\textit{Phys. Rev. Lett.} \textbf{102}, 086803 (2009).

\bibitem{BergesPRL2004} J. Berges, S. Bors\'{a}nyi, and C. Wetterich, 
\textit{Phys. Rev. Lett.} \textbf{93}, 142002 (2004).

\bibitem{JiNature2003} Y. Ji, Y. Chung, D. Sprinzak, M. Heiblum, D. Mahalu,
and H. Shtrikman, \textit{Nature} \textbf{422}, 415 (2003).

\bibitem{LevkivskyiPRB2008} I. P. Levkivskyi and E. V. Sukhorukov, \textit{%
Phys. Rev. B} \textbf{78}, 045322 (2008).

\bibitem{ProkudinaPRB2010} M. G. Prokudina, V. S. Khrapai, S. Ludwig, J. P.
Kotthaus, H. P. Tranitz, and W. Wegscheider,\textit{\ Phys. Rev. B} \textbf{%
82}, 201310 (2010).

\bibitem{LundePRB2010} A. M. Lunde, S. E. Nigg, and M. B\"{u}ttiker, \textit{%
Phys. Rev. B} \textbf{81}, 041311 (2010).
\end{thebibliography}
\end{document}